\documentstyle[sprocl,epsf,epsfig]{article}
 \catcode`@=11
\def\@cite#1#2{\unskip\nobreak\relax
    \def\@tempa{$\m@th^{\hbox{\the\scriptfont0 #1}}$}%
    \futurelet\@tempc\@citexx}
\def\@citexx{\ifx.\@tempc\let\@tempd=\@citepunct\else
    \ifx,\@tempc\let\@tempd=\@citepunct\else
    \let\@tempd=\@tempa\fi\fi\@tempd}
\def\@citepunct{\@tempc\edef\@sf{\spacefactor=\the\spacefactor\relax}\@tempa
    \@sf\@gobble}

\def\citenum#1{{\def\@cite##1##2{##1}\cite{#1}}}
\def\citea#1{\@cite{#1}{}}

\newcount\@tempcntc
\def\@citex[#1]#2{\if@filesw\immediate\write\@auxout{\string\citation{#2}}\fi
  \@tempcnta\z@\@tempcntb\m@ne\def\@citea{}\@cite{\@for\@citeb:=#2\do
    {\@ifundefined
       {b@\@citeb}{\@citeo\@tempcntb\m@ne\@citea\def\@citea{,}{\bf ?}\@warning
       {Citation `\@citeb' on page \thepage \space undefined}}%
    {\setbox\z@\hbox{\global\@tempcntc0\csname b@\@citeb\endcsname\relax}%
     \ifnum\@tempcntc=\z@ \@citeo\@tempcntb\m@ne
       \@citea\def\@citea{,}\hbox{\csname b@\@citeb\endcsname}%
     \else
      \advance\@tempcntb\@ne
      \ifnum\@tempcntb=\@tempcntc
      \else\advance\@tempcntb\m@ne\@citeo
      \@tempcnta\@tempcntc\@tempcntb\@tempcntc\fi\fi}}\@citeo}{#1}}
\def\@citeo{\ifnum\@tempcnta>\@tempcntb\else\@citea\def\@citea{,}%
  \ifnum\@tempcnta=\@tempcntb\the\@tempcnta\else
   {\advance\@tempcnta\@ne\ifnum\@tempcnta=\@tempcntb \else \def\@citea{--}\fi
    \advance\@tempcnta\m@ne\the\@tempcnta\@citea\the\@tempcntb}\fi\fi}
\catcode`@=12

\renewenvironment{thebibliography}[1]
 {\begin{list}{\arabic{enumi}.}
    {\usecounter{enumi} \setlength{\parsep}{0pt}
     \setlength{\itemsep}{3pt} \settowidth{\labelwidth}{#1.}
     \sloppy
    }}{\end{list}}

\begin{document}

\vspace*{-.4in}

\font\fortssbx=cmssbx10 scaled \magstep1
\hbox to \hsize{
\includegraphics{uwlogo.ps}
\hskip.25in \raise.1in\hbox{\fortssbx University of Wisconsin - Madison}
\hfill$\vcenter{\hbox{\bf MADPH-96-965}
                \hbox{October 1996}}$ }

\vskip.2in

\title{
\uppercase{(No) Color in QCD: Charmonium, Charm and Rapidity Gaps}%
\,\footnote{Talk presented by F. Halzen at the {\it 26th International
Symposium on Multiparticle Dynamics (ISMD\,96)}, Faro, Portugal, 1996}}

\author{\vskip-.2in
    \uppercase{O. J. P. \'Eboli and E. M. Gregores}}
\address{Instituto de
    F\'{\i}sica, Universidade de S\~ao Paulo,
     S\~ao Paulo, Brazil}
\author{\uppercase{F. Halzen}}
\address{Department of Physics, University of Wisconsin, Madison, WI
  53706, USA}

\maketitle
\abstracts{
The reason why some data on the production of  $\psi$- and $\Upsilon$-states
disagree with QCD predictions, occasionally by well over one order of
magnitude\cite{review}, is that the traditional method for performing the
perturbative calculation of the cross section is simply wrong. The key mistake
is to require that the heavy quark pair forms a color singlet at short
distances, given that there is an infinite time for soft gluons to readjust the
color of the $c \bar c$ pair before it appears as an asymptotic $\psi$ or,
alternatively, $D \bar D$ state. We suspect that the same mistake is made in
the description of rapidity gaps, i.e.\ the production of a color-neutral
quark-antiquark pair, in terms of the exchange of a color neutral gluon pair.
The $\psi$ is after all a color neutral $c \bar c$ pair and we will show that
it is produced by the same dynamics as $D \bar D$ pairs; its color happens to
be bleached by soft final-state interactions. This approach to color is
suggestive of the unorthodox prescription for the production of rapidity gaps
in deep inelastic scattering, proposed by Buchm\"uller and Hebecker\cite{bh}.
When applied to the formation of gaps between a pair of high transverse
momentum jets in hadron collisions, the soft color approach suggests a
formation rate of gaps in gluon-gluon subprocesses which is similar or smaller
than in quark-quark induced events. Formation of gaps should increase when
increasing transverse momentum or lowering energy, in contrast with 2-gluon
exchange Pomeron models.}

\section{Introduction}
\unskip\smallskip
The conventional treatment of color, i.e., the color singlet model, has run
into serious problems describing the data on the production of charmonium and
upsilon states\cite{review}. Specific proposals to solve the charmonium problem
agree on the basic solution: its production is a two-step process where a heavy
quark pair is produced first. At this stage perturbative diagrams are included
whether the $c \bar c$ pair is color singlet or not. This is a departure of the
textbook approach where only diagrams where the charm pair is in a color
singlet are selected. In the Bodwin-Braaten-Lepage (BBL) formalism\cite{bbl}
the subsequent evolution of the pair into a colorless bound state is described
by an expansion in powers of the relative velocity of the heavy quarks in the
onium system. An alternative approach, color evaporation or the soft color
method, represents an even more radical departure from the way color singlet
states are conventionally treated in perturbation theory.  Color is, in fact,
``ignored''.  Rather than explicitly imposing that the system is in a color
singlet state in the short-distance perturbative diagrams, the appearance of
color singlet asymptotic states depends solely on the outcome of large-distance
fluctuations of quarks and gluons. In other words, color is a nonperturbative
phenomenon.

\begin{figure}[h]
\begin{center}
        \begin{tabular}{ccc}
                \epsfxsize=0.22\hsize
                \mbox{\epsffile{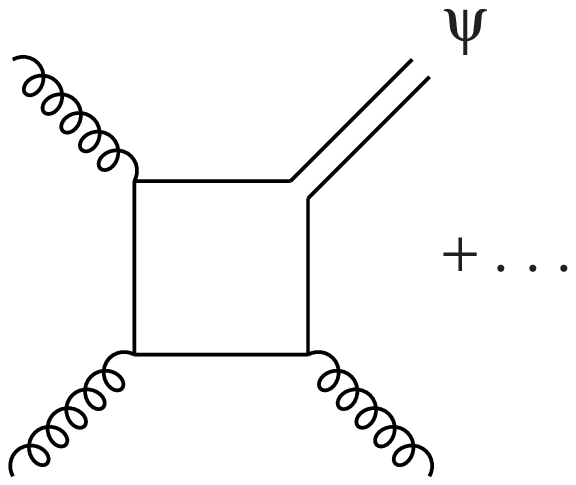}}
                &~&
                \epsfxsize=0.33\hsize
                \mbox{\epsffile{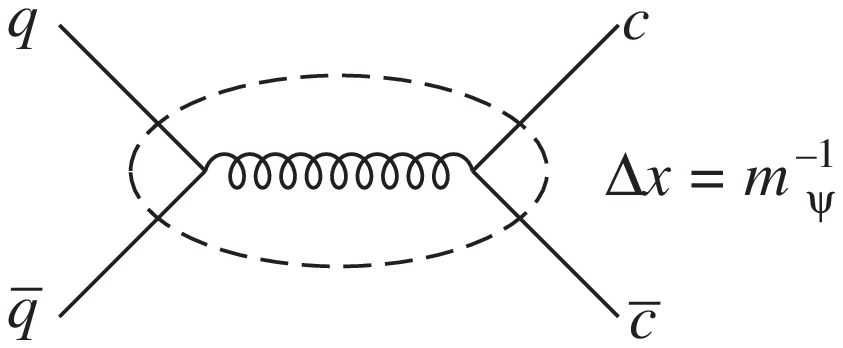}}
                \\
                (a)
                &~&
                (b)
                \\
        \end{tabular}
\end{center}

\vskip-.2in
\caption{Typical diagrams for (a) color singlet $\psi$ production and (b)
color evaporation $\psi$ production.
\label{fig:csm}
\label{fig:cem}}
\end{figure}

In Fig.~\ref{fig:csm} we show typical diagrams for the production of
$\psi$-particles representing the competing treatments of the color quantum
number. In the diagram of Fig.~\ref{fig:csm}a, the color singlet approach, the
$\psi$ is produced in gluon-gluon interactions in association with a final
state gluon which is required by color conservation. This diagram is related by
crossing to the hadronic decay $\psi \rightarrow 3$ gluons. In the color
evaporation approach, the color singlet property of the $\psi$ is ignored at
the perturbative stage of the calculation. The $\psi$ can, for instance, be
produced to leading order by $q\bar q$-annihilation into $c\bar c$, which is
the color-equivalent of the Drell-Yan process.  This diagram is calculated
perturbatively; its dynamics are dictated by short-distance interactions of
range $\Delta x \simeq m_{\psi}^{-1}$. It does indeed not
seem logical to enforce the color singlet property of the $\psi$ at short
distances, given that there is an infinite time for soft gluons to readjust the
color of the $c \bar c$ pair before it appears as an asymptotic $\psi$ or,
alternatively, $D \bar D$ state. Alternatively, it is indeed
hard to imagine that a color singlet state formed at a range $m_{\psi}^{-1}$,
automatically survives to form a $\psi$. This formalism represents the original
and, as we will show, correct method by which perturbative QCD calculations
were performed\cite{cem,fh:1a,fh:1b,gor}.

We will first discuss the resolution of the charmonium problem, emphasizing the
color evaporation approach. The solution suggests a radical departure from the
way color is treated in perturbative QCD calculations. We will subsequently
speculate on the implications for the dynamics underlying the production of
rapidity gaps which refer to regions in phase space where no hadrons appears as
a result of the production of a color neutral quark-antiquark pair. The
connection to charmonium physics is obvious: the $\psi$ is a color-neutral $c
\bar c$~pair!

\section{Onium Calculations with Soft Color}

The ``color evaporation" or ``soft color" treatment of the color quantum number
lead to a similar description of bound and open charm production:
\begin{equation}
\sigma_{\rm onium} = \frac{1}{9} \int_{2 m_c}^{2 m_D} dm~
\frac{d \sigma_{c \bar{c}}}{dm} \; ,
\label{sig:on}
\end{equation}
and
\begin{eqnarray}
\sigma_{\rm open} &=& \frac{8}{9}  \int_{2 m_c}^{2 m_D} dm~
\frac{d \sigma_{c \bar{c}}}{dm}
+ \int_{2 m_D} dm~\frac{d \sigma_{c \bar{c}}}{dm}
\label{sig:op}
\\
& \simeq &\frac{8}{9}  \int_{2 m_c} dm~
\frac{d \sigma_{c \bar{c}}}{dm}
\label{sig:ap}
\end{eqnarray}
where the cross section for producing heavy quarks, $\sigma_{c \bar c}$, is
computed perturbatively.  Diagrams are included order-by-order, irrespective of
the color of the $c \bar c$ pair. The coefficients $\frac{1}{9}$ and
$\frac{8}{9}$ represent the statistical probabilities that the $3\times\bar3$
charm pair is asymptotically in a singlet or octet state. In order to achieve
the phenomenological success described here it is essential to systematically
include next-to-leading order terms. Neglecting O($\alpha_s^3$) terms is
equivalent to describing photon interactions with matter neglecting the
Bethe-Heitler process versus Compton scattering because it is a higher order
process. The former actually dominates at high energy for reasons that are
similar to those requiring the inclusion of higher order heavy quark processes.

\begin{figure}[t]
\begin{center}
\mbox{\epsfig{file=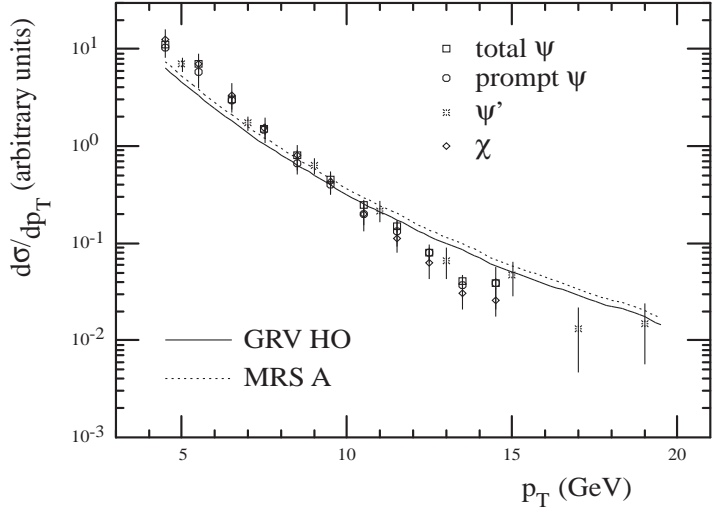,width=.57\linewidth,angle=-90}}
\end{center}

\caption{Data from the CDF Collaboration \protect\cite{exp:pt}, shown
with arbitrary normalization.  The curves are the predictions of the
color evaporation model at tree level, also shown with arbitrary
normalization. The normalization is correctly predicted within a
K-factor of 2.2.\label{fig:cdf-pt}}
\end{figure}

\begin{figure}[t]
\begin{center}
\mbox{\epsfig{file=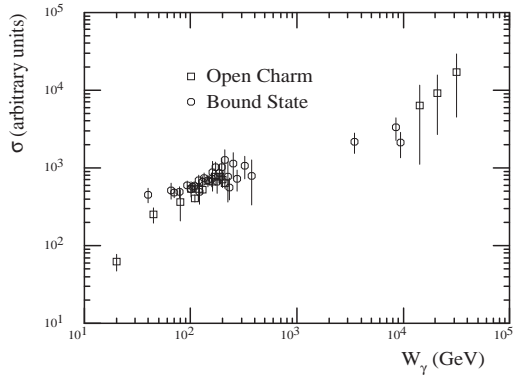,angle=-90,width=.57\linewidth}}
\end{center}

\caption{Photoproduction data\protect\cite{data1,data2} as a function
  of the photon energy in the hadron rest frame, $W_\gamma$.  The
  normalization has been adjusted to show the similar shapes of the
  data.
\label{fig:justdata}}
\end{figure}

\begin{figure}[t]
\begin{center}
\mbox{\epsfxsize=1.5in\epsfbox{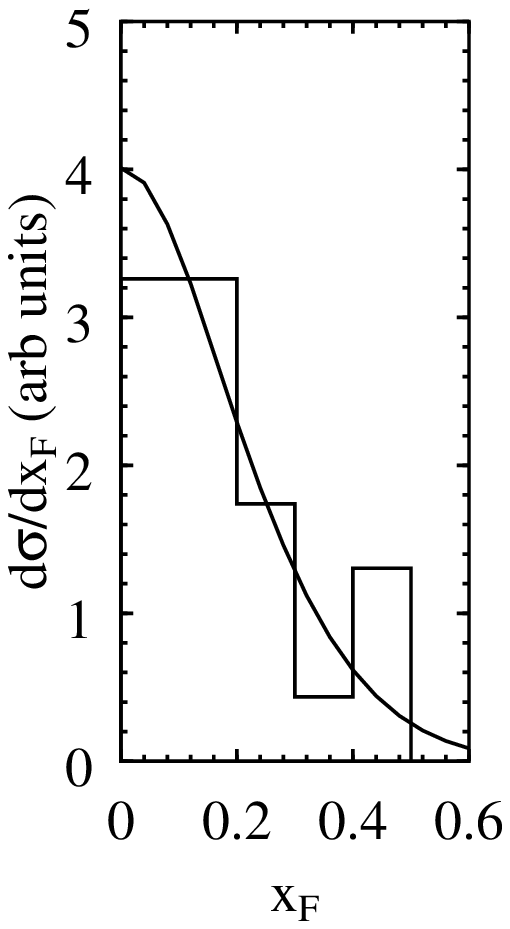}
\epsfxsize=1.5in\epsfbox{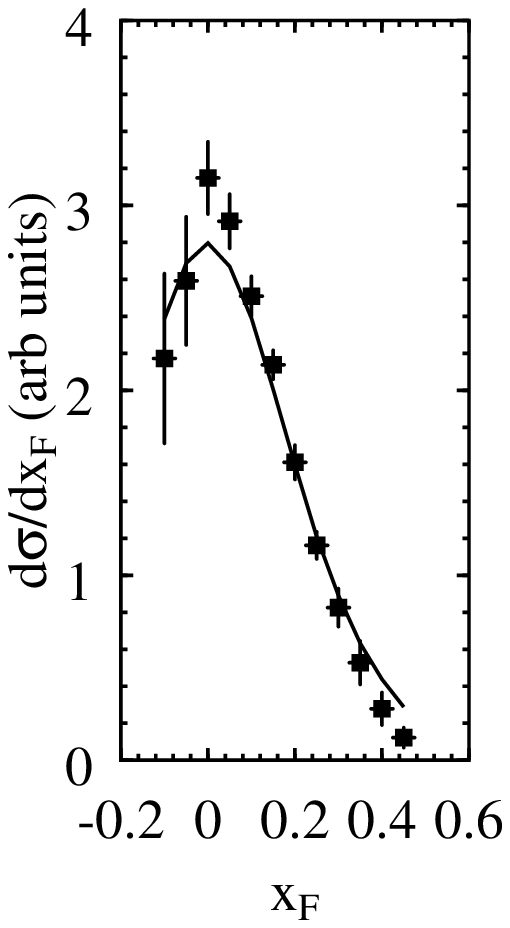}}
\end{center}
\vskip-.6in
\caption{ Normalized $x_F$
distribution for the production of of $D\bar{D}$ (histogram) at
$\protect\sqrt{s} = 27.4$ GeV and $J/\psi$ (squares) in proton-nucleon
collisions at $\protect\sqrt{s} = 23.7$ GeV. The curves are the
prediction of our model.\label{fig:xf}}
\end{figure}

In principal the calculation only predicts the sum of the cross sections of all
onium states given by Eq.~(\ref{sig:on}). This sum rule is, unfortunately,
difficult to test experimentally, since it requires measuring cross sections
for {\em all} of the bound states at a given energy. This does not mean that
the calculation has no predictive power. The above equations make the bold
prediction that all onium states $\psi$, $\psi$', $\chi$ and $\eta_c$ states
share the same production dynamics which they also share with open charm in the
limit $m_c \simeq m_D$; see Eq.~(3). The CDF collaboration has accumulated
large samples of data on the production of prompt $\psi$, $\chi_{cJ}$, and
$\psi^\prime$\cite{exp:pt}. Since all charmonium states share the same
production dynamics in the color evaporation scheme, their $p_T$ distributions
should be the same, up to a multiplicative constant.  This prediction is borne
out by the CDF data, as we can see in Fig.~\ref{fig:cdf-pt}. We will return to
a detailed calculation of the distribution shown further on. The formalism
predicts furthermore that, up to color and normalization factors, the energy,
$x_F$ and $p_T$ dependence of production cross sections for onium states and
open charm pairs is the same. Support for the prediction of Eqs.~(\ref{sig:on})
and (\ref{sig:ap}) that the production of hidden and open charm have similar
dynamics is shown in Fig.~\ref{fig:justdata}, which displays charm
photoproduction data for both open charm and bound state production with common
normalization in order to show their identical energy behavior.  A similar
figure for hadroproduction can be found in Ref.~11. By the same argument the
formalism also predicts that the normalized $x_F$ distribution for $J/\psi$ and
$D\bar{D}$ pairs should be the same at a given center-of-mass energy. This is
indeed the case \cite{exp:xfddbar,exp:xfpsi}; see Fig.~\ref{fig:xf}.

\begin{figure}[t]
\begin{center}
\vspace{.5in}
\mbox{\epsfig{file=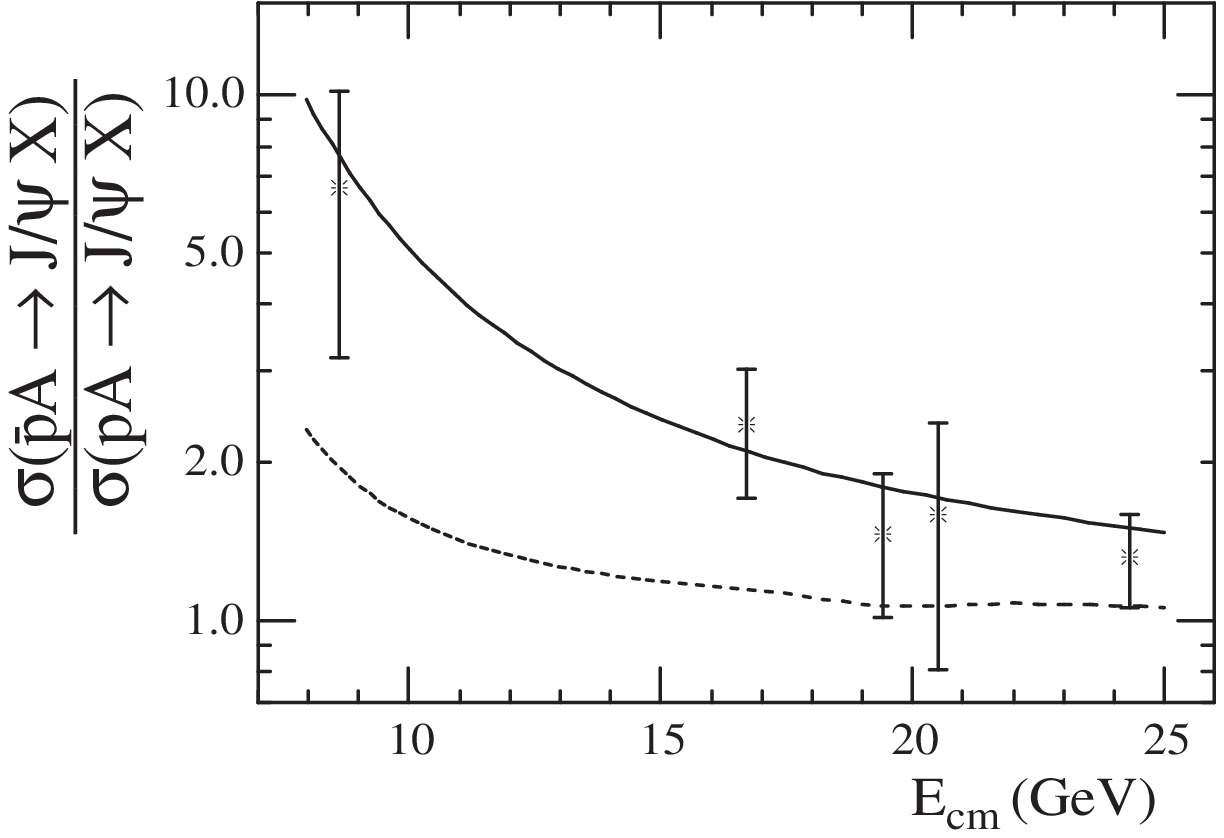,width=.6\linewidth,angle=-90}}
\end{center}

\caption{Ratio of the cross sections for the production of $J/\psi$
by proton and antiproton beams in the color evaporation model (solid
line) and the color singlet model (dashed line) as a function of the
center-of-mass energy.  Data taken from
Ref.~14.
\label{fig:part-anti}}
\end{figure}

\break
One of the most striking features of color evaporation is that the production
of charmonium is dominated by the conversion of a colored gluon into a $\psi$,
as in Fig.~\ref{fig:cem}b.  In the
conventional treatment, where color singlet states are formed at the
perturbative level, 3 gluons (or 2 gluons and a photon) are required to produce
a $\psi$. Contrary to the usual folklore, $\psi$'s are, except at the higher
energies, not produced by gluons. As a consequence color evaporation predicts
an enhanced $\psi$ cross section for antiproton beams, while the color singlet
model predicts roughly equal cross sections for proton and antiproton beams.
The prediction of an enhanced $\bar p$ yield is obviously correct: antiproton
production of  $\psi$'s exceeds that by protons by a factor 5 close to
threshold; see Fig.~\ref{fig:part-anti}. This fact has been known for some
time\cite{fh:1a,fh:1b,gor}. We should note that for sufficiently high energies,
gluon initial states will eventually dominate because they represent the bulk
of soft partons.

\section{Quantitative Tests of Soft Color}

The color evaporation scheme assumes a factorization of the production of the
$c\bar{c}$ pair, which is perturbative and process dependent, and the
materialization of this pair into a charmonium state by a mechanism that is
nonperturbative and process independent. This
assumption is reasonable since the characteristic time scales of the two
processes are very different: the time scale for the production of the pair is
the inverse of the heavy quark mass, while the formation of the bound state is
longer than the time scale $1/\Lambda_{\rm QCD}$.  Therefore, explicit
comparison with the $\psi$ data requires knowledge of the fraction $\rho_\psi$
of produced onium states that materialize as $\psi$'s, {\em i.e.,}
\begin{equation}
\sigma_\psi = \rho_\psi \sigma_{\rm onium} \; ,
\label{frac}
\end{equation}
where $\rho_\psi$ is assumed to be a constant, independent of the process. This
assumption is in agreement with the low energy data \cite{gksssv,schuler}. The
constant not only accounts for the direct production of $\psi$ to the onium
cross section, but also includes its production via $\psi'$ and $\chi$
production and decay.

Quantitative tests of color evaporation are made possible by the fact that all
$\psi$-production data, i.e.\ photo-, hadroproduction, $Z$-decay, etc., are
described in terms of a single parameter. Once $\rho_\psi$ has been empirically
determined for one initial state, the cross section is predicted without free
parameters for the other.  We will illustrate the power of the color
evaporation scheme by showing how it quantitatively accommodates all
measurements, including the high energy Tevatron and HERA data, which have
represented a considerable challenge for the color singlet model. Its
parameter-free prediction for the  $Z$-boson decay rate into $\psi$'s is an
order of magnitude larger than the color singlet model and consistent with
data.

\begin{figure}[h]
\begin{center}
\mbox{\epsfig{file=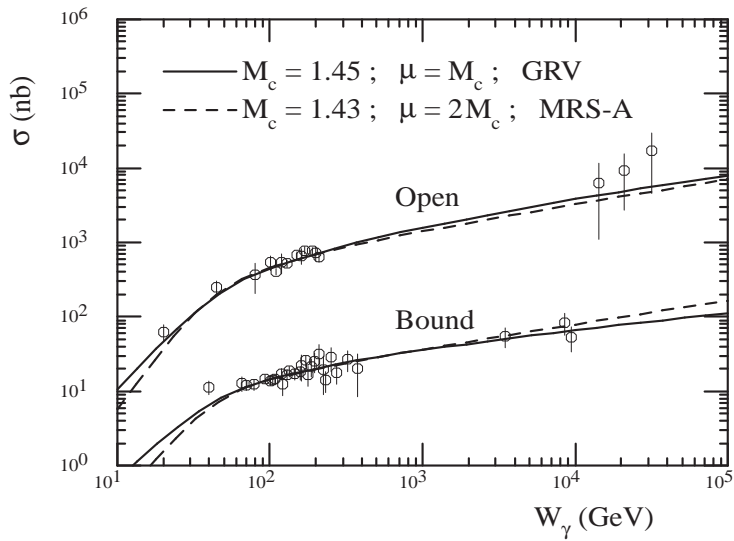,width=.6\linewidth,angle=-90}}
\end{center}

\caption{
  Photoproduction data \protect\cite{data1,data2} and the predictions
  of the color evaporation model at next-to-leading order as a
  function of the photon energy in the hadron rest frame, $W_\gamma$.
  The normalizations in this figure are absolute.
\label{fig:photopro}}
\end{figure}

In Fig.~\ref{fig:photopro} we compare the photoproduction data with theory,
using the NLO perturbative QCD calculation of charm pair production from
Ref.~17. From the relative magnitude of the $\psi$ and open charm cross
sections we determine the fragmentation factor $\rho_\psi$ to be 0.50 using GRV
HO, or 0.43 using MRS~A structure functions. Note that the factor $\rho_\psi$
possesses a theoretical uncertainty due to the choice of scales and parton
distribution functions.  We conclude the photoproduction of $J/\psi$ and
$D\bar{D}$ is well described by the color evaporation model.  This reaction has
now been used to fix the only free parameter, $\rho_\psi\approx 0.5$.

\begin{figure}
\begin{center}
\mbox{\epsfig{file=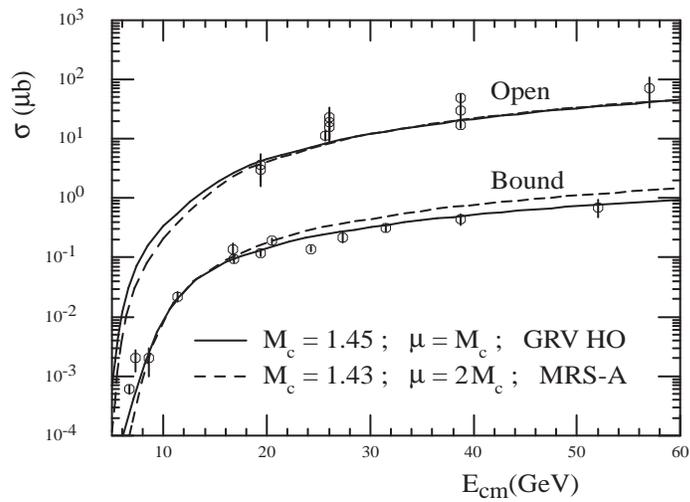,width=.6\linewidth,angle=-90}}
\end{center}

\caption{
  Hadroproduction data \protect\cite{exp:psi,exp:charm} and the
  predictions of the color evaporation model at next-to-leading order
  as a function of the center-of-mass energy, $E_{cm}$.  The curve
  for bound state production is an absolutely normalized,
  parameter-free prediction of the color evaporation
model.\label{fig:hadro-fit}}
\end{figure}

At this point the predictions of the color evaporation model for
hadroproduction of $\psi$ are completely determined, up to O($\alpha_s^4$) QCD
corrections. In Fig.~\ref{fig:hadro-fit} we compare the color evaporation model
predictions with the data. and conclude that the this color scheme describes
the hadroproduction very accurately. In order to also obtain a theoretical
prediction for the $p_T$-distribution already shown in Fig.~\ref{fig:cdf-pt},
we have computed the processes $g + g \to [c\bar{c}] + g$, $q + \bar{q} \to
[c\bar{c}] + g$, and $g + q \to [c\bar{c}] + q$ at tree level using MADGRAPH
\cite{tim}.  We imposed that the $c\bar{c}$ pair satisfy the invariant mass
constraint of Eq.~(\ref{sig:on}).  Our results are shown in
Fig.~\ref{fig:cdf-pt}. Higher order corrections such as soft-gluon resummation
are expected to tilt our lowest order prediction, bringing it to a closer
agreement with the data\cite{fleming}.

In the color-evaporation scheme the width for inclusive $Z$ decay into prompt
charmonium is:
\begin{equation}
\Gamma( Z \rightarrow \mbox{prompt charmonium}) =
\frac{1}{9} \int_{2 m_c}^{2 m_D} dm~
\frac{d \hat{\Gamma}_{c \bar{c}}}{dm} \; ,
\end{equation}
where $\hat{\Gamma}$ is the partonic width for producing a $c \bar{c}$ pair.
The procedure should by now be familiar: in order to obtain the partial width
into a specific charmonium state we multiply the above expression by the
appropriate fragmentation fraction $\rho$ into $\psi$, which was determined
from charmonium photoproduction data.  Notice that the predictions
for the $Z$ decay into charmonium are parameter-free.

We have again evaluated all the tree-level partonic amplitudes using the
package MADGRAPH \cite{tim}. Although formally of higher order in $\alpha_s$,
the dominant process for the inclusive decay of the $Z$ into charmonium is $Z
\rightarrow c \bar{c} q
\bar{q}$, where $q=u$, $d$, $s$, $c$, and $b$. (The leading-order process in
$\alpha_s$ is $Z \rightarrow c\bar{c} g$, which leads to the production of a
charmonium state and a hard
jet is suppressed by a virtual quark propagator of order $m_c/m_Z$). The
branching fraction of $Z$ into prompt $\psi$ is
$(1.7\mbox{--}1.8)\times10^{-4}$. This is to be contrasted
with the color-singlet model which predicts a branching fraction for direct
$\psi$ in $Z$ decay of the order $3\times10^{-5}$ \cite{bcy}. The
color-evaporation model leads to a branching fraction larger by almost an order
of magnitude consistent with the result reported by the OPAL collaboration of
\[
{\rm B}( Z \rightarrow {\rm prompt}~ \psi + X) =
(1.9 \pm 0.7 \pm 0.5 \pm 0.5) \times 10^{-4} \; .
\]

We hope that we have illustrated by now that the soft color approach gives a
complete picture of charmonium production in hadron-hadron, $\gamma$-hadron,
and $Z$ decays. The phenomenological success of the soft color scheme is
impressive and extends to applications to other charmonium and upsilon
states\cite{amundson,schuler}.

\section{Intermezzo: Soft Color and BBL}

Other approaches, very similar in spirit, can be found in Refs.~3, 15 and 23.
The color evaporation approach differs from Ref.~3, the formalism of Bodwin,
Braaten and Lepage, in the way that the $c \bar c$ pair exchanges color with
the
underlying event.  In the BBL formalism, multiple gluon interactions with the
$c\bar c$ pair are suppressed by powers of $v$, the relative velocity of the
heavy quarks within the $\psi$. The color evaporation model assumes that these
low-energy interactions can take place through multiple, soft-gluon
interactions. While the formalism allows straightforward application to heavy
quark decays, it is not always clear how to compute production cross sections
in the BBL formalism (e.g.\ photoproduction of $\psi$ near
$z=1$)\cite{fleming}. The color evaporation scheme, though partly
nonperturbative, is phenomenologically well-defined, has less parameters (1
versus 3 for describing $\psi$-production). Also, next-to-leading order
corrections are included in a straightforward way, a necessary condition for
obtaining quantitative predictions.

The $\psi$'s produced through the color-evaporation mechanism are expected to
be unpolarized since the polarization information is lost because of the
multiple soft gluon exchanges \cite{mirkes}. On the other hand, the
(non)polarization of $\psi$ is hard to explain in the framework of the
color-octet model\cite{pol:com,braatenchen}. Therefore, the measurement of the
polarization of the produced charmonium may very well be a tool to discriminate
between these competing descriptions.

\section{Implications for the Physics of Rapidity Gaps}

\begin{figure}
\begin{center}
\leavevmode\epsfxsize=1.75in\epsfbox{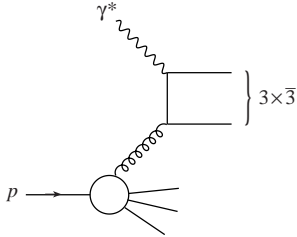}
\end{center}
\caption{
Mechanism for the production of rapidity gaps in
deep inelastic scattering.}
\label{gap:dis}
\end{figure}

The important lesson about color resides however in the similarity, not the
differences of these approaches: perturbative color octet states fully
contribute to the asymptotic production of color singlet states such as
$\psi$'s. We suspect that this is also true for the production of a rapidity
gap which is, e.g.\ when produced in electroproduction, nothing but the
creation of a color singlet quark-antiquark pair; see Fig.~\ref{gap:dis}. The
diagram shown represents the production of  final state hadrons which are
ordered in rapidity.  From top to bottom we find the fragments of the
intermediate partonic quark-antiquark state and those of the target.
Buchm\"uller and Hebecker proposed that the origin of a rapidity gap
corresponds to the absence of color between photon and proton, {\em i.e.}  the
$\mbox{\bf 3} \times \bar{\mbox{\bf 3}}$ ($= \mbox{\bf 1} + \mbox{\bf 8}$)
intermediate quark-antiquark state is in a color singlet state.  Because color
is the source of hadrons, only the color octet states yield hadronic asymptotic
states.  This leads to the approximate expectation that
\begin{equation}
F_2^{(gap)} = \frac{1}{1+8} F_2
\end{equation}
Although this result is subject to corrections, it embodies the essential
physics: events with and without gaps are described by the same short-distance
dynamics. Essentially non-perturbative final-state interactions dictate the
appearance of gaps whose frequency is determined by simple counting. The
treatment of color is the same as in the case of heavy quark production: the
same perturbative mechanisms, i.e.\ gluon exchange, dictates the dynamics of
color-singlet gap ($\psi$) and regular deep inelastic (open charm) events.

\begin{figure}
\begin{center}
\leavevmode\epsfxsize=1.75in\epsfbox{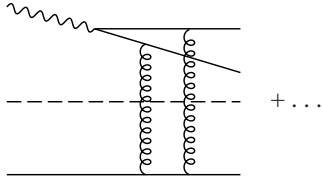}
\end{center}
\caption{
Pomeron mechanism for
the formation of rapidity gaps.}
\label{gap:pom}
\end{figure}

\begin{figure}[h]
\begin{center}
\leavevmode\epsfxsize=1.75in\epsfbox{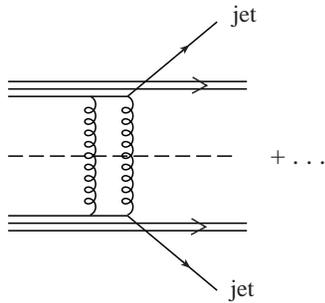}
\end{center}
\caption{
Pomeron mechanism for
the formation of rapidity gaps in hadron collisions.}
\label{2j:pom}
\end{figure}

\begin{figure}[t]
\begin{center}
\leavevmode\epsfxsize=1.75in\epsfbox{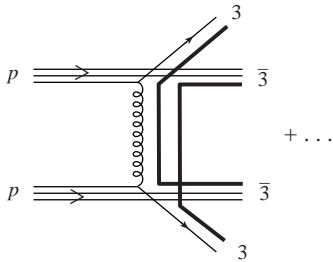}
\end{center}
\caption{
Color bleaching  picture for
the formation of rapidity gaps in hadron collisions.}
\label{2j:ble}
\end{figure}

Our understanding of the (soft) nature of color challanges the orthodox
description of rapidity gaps in terms of the so-called hard Pomeron description
sketched in Fig.~\ref{gap:pom}. The $t$-channel exchange of a pair of gluons in
a color singlet state is the origin of the gap. The color string which connects
photon and proton in diagrams such as the one in Fig.~\ref{gap:dis}, is absent
and no hadrons are produced in the rapidity region separating them.  The same
mechanism predicts rapidity gaps between a pair of jets produced in hadronic
collisions; see Fig.~\ref{2j:pom}. These have been observed and occur with a
frequency of
order of one percent~\cite{d0}. The arguments developed in this work invalidate
this approach: it is as meaningless to enforce the color singlet nature of the
gluon pair as it is to require that the $c \bar c$ pair producing $\psi$ is
colorless at the perturbative level. Following our color scheme the gaps are
accommodated as a mere final state color bleaching phenomenon {\em\`{a} la}
Buchm\"uller and Hebecker. This can be visualized using the diagram shown in
Fig.~\ref{2j:ble}. At short distances it represents a conventional perturbative
diagram for the production of a pair of jets. Also shown is the string picture
for the formation of the final state hadrons. Color in the final state is
bleached by strings connecting the ${\mbox{\bf 3}}$ jet at the top with the
$\bar{\mbox{\bf 3}}$ spectator di-quark at the bottom and vice-versa. The
probability to form a gap can be counted {\em \`{a} la} Buchm\"uller and
Hebecker to be $1/(1+8)^2$ because it requires the formation of singlets in 2
strings. This is consistent with observation and predicts that, as was the case
for electroproduction, the same short distance dynamics governs events with and
without rapidity gaps.  The data \cite{d0} is consistent with the prediction of
this simple picture which basically predicts that the gap fraction between $pp$
jets is the square of that between virtual photon and proton in deep inelastic
scattering.

One should realize that this string picture is not necessarily the correct one.
It is more likely that the color is bleached between the top and bottom
${\mbox{\bf 3}}$ and $\bar{\mbox{\bf 3}}$ which are widely separated in
rapidity space.

This discussion ignores that gluon-gluon as well as quark-quark subprocesses
contribute to jet production in hadron collisions. In the color flow diagram
corresponding to Fig.~\ref{2j:ble} top and bottom protons each split into a
color octet gluon and color octet 3-quark remnant. There are now $(8 \times
8)^2$ color final states. We anticipate a reduced probability to form a color
singlet. The reduction may not be very significant because the $10 +
\overline{10}$ and 27 color final states may be suppressed. One argument for
this is that these representations consist of exotic multi-quark states which
do not materialize into final state mesons. High color charges may also be
suppressed for dynamical reasons. Despite the fact that we can at best guess
the non-perturbative dynamics, it is clear that the soft color formalism
predicts a gap rate which is similar of smaller in gluon-gluon interactions.
This is in contrast with the diagram of Fig.~\ref{gap:pom} which predicts a
gap-rate enhanced by a factor $\left(9 \over 4\right)^2$ in gluon-gluon
subprocesses\cite{zeppo}. The contrasting predictions can be easily tested by
enhancing the relative importance of quark-quark subprocesses. i.e. by
increasing the $p_T$ of the jets at fixed energy of by decreasing the collision
energy of the hadrons at fixed $p_T$. In either case we anticipate in the soft
color scheme an increased rate for the production of gaps, a prediction
opposite from that expected in the 2-gluon exchange model\cite{hiz}.

Do Figs.~\ref{gap:dis}--\ref{2j:ble} suggest that we have formulated
alternative $s$- and $t$-channel pictures to view the same physics? Although
they seem at first radically different, this may not be the case. Computation
of the exchange of a pair of colorless gluons in the $t$-channel is not
straightforward and embodies all the unsolved mysteries of constructing the
``Pomeron'' in QCD. In a class of models where the Pomeron is constructed out
of gluons with a dynamically generated mass \cite{natale,chehime}, the diagram
of Fig.~\ref{2j:pom} is, not surprisingly, dominated by the configuration where
one gluon is hard and the other soft. The diagram is identical to the standard
perturbative diagram except for the presence of a soft, long-wavelength gluon
whose only role is to bleach color. Its dynamical role is minimal, events with
gaps are not really different from events without them. Soft gluons readjust
the color at large distances and long times. Their description is outside the
realm of perturbative QCD. In this class of models the hard Pomeron is expected
to be no more than an order $\alpha_s^2$ correction, a view which can be
defended on more solid theoretical ground \cite{cudell}.

Some have challenged the theoretical soundness of this line of
thinking\cite{bj,white}. Also note that our discussion is at best indirectly
relevant to completely non-perturbative phenomena like elastic scattering.
There is no short distance limit defined by a large scale.  The Pomeron exists.

\section*{Acknowledgments}

We would like to thank J.~Amundson for collaborations and J.~Bjorken,
G.~ingelman, A.~White, D~Zeppenfeld and S.~Fleming for his insight.  This
research was supported in part by the University of Wisconsin Research
Committee with funds granted by the Wisconsin Alumni Research Foundation, by
the U.S.\ Department of Energy under grant DE-FG02-95ER40896, and by Conselho
Nacional de Desenvolvimento
Cient\'{\i}fico e Tecnol\'ogico (CNPq).

\end{document}